\documentclass[reprint
superscriptaddress,
showpacs,
 amsmath,amssymb,prl
]{revtex4-2}
\usepackage{dcolumn}
\usepackage{bm}
\newtheorem{theorem}{Theorem}
\newtheorem{corollary}{Corollary}[theorem]

\newtheorem{definition}{Definition}

\begin{document}
\title{An advance in the arithmetic of the Lie groups as an alternative to the forms of the Campbell-Baker-Hausdorff-Dynkin theorem}
\author{Sunghyun Kim}
\email{Sunghyun.Kim@ucf.edu}
\affiliation{Department of Physics, University of Central Florida, Orlando, FL 32816-2385, USA}
\author{Zhichen Liu}
\email{Zhichen.Liu@ucf.edu}
\affiliation{Department of Physics, University of Central Florida, Orlando, FL 32816-2385, USA}
\author{Richard A. Klemm}
\email{richard.klemm@ucf.edu, corresponding author}
\affiliation{Department of Physics, University of Central Florida, Orlando, FL 32816-2385, USA}
\affiliation{U. S. Air Force Research Laboratory, Wright-Patterson Air Force Base, Ohio 45433-7251, USA}
\date{\today}

\begin{abstract}

 The exponential of an operator or matrix is widely used in quantum theory, but it sometimes can be a challenge to evaluate. For non-commutative operators ${\bf X}$ and ${\bf Y}$, according to the Campbell-Baker-Hausdorff-Dynkin theorem, ${\rm  e}^{{\bf X}+{\bf Y}}$ is not equivalent to ${\rm e}^{\bf X}{\rm e}^{\bf Y}$, but is instead given by the well-known infinite series formula. For a Lie algebra of a basis of three operators $\{{\bf X,Y,Z}\}$, such that $[{\bf X}, {\bf Y}] = \kappa{\bf Z}$ for scalar $\kappa$ and cyclic permutations, here it is proven that ${\rm e}^{a{\bf X}+b{\bf Y}}$ is equivalent to ${\rm e}^{p{\bf Z}}{\rm e}^{q{\bf X}}{\rm e}^{-p{\bf Z}}$ for scalar $p$ and $q$. Extensions for ${\rm e}^{a{\bf X}+b{\bf Y}+c{\bf Z}}$ are also provided. This method is useful for the dynamics of atomic and molecular nuclear and electronic spins in  constant and oscillatory transverse magnetic and electric fields.\\
\end{abstract}

 \vskip0pt
\maketitle

\section{Introduction}

In 1954, Rabi, Ramsey and Schwinger reviewed magnetic resonance problems in the rotating coordinates of a nuclear spin \cite{rabi}. The reviewed literature focused on the spin in a weak rotating magnetic field normal to a strong constant magnetic field and the transition probability from a singly occupied state to another singly occupied state, rather than upon the quantum spin wave function. A brief derivation of the spin wave function  that satisfies the Schrödinger equation at the time $t$ was later found  by Gottfried to be
    \begin{eqnarray} \label{gottfriedfunction}
        \Psi(t) &=&
        {\rm e}^{{\rm i}\omega t {\bm J}_{z}}
       {\rm e}^{-{\rm i}[({\omega-\Omega}){\bm J}_{z} - \lambda\Omega {\bm J}_{x}]t}
        \Psi(0)
    \end{eqnarray}
where $\omega,\Omega$ are scalar frequencies, $\lambda$ is a small dimensionless parameter, and ${\bm J}_{x}, {\bm J}_{z}$ are spin operators  in units of $\hbar=h/(2\pi)$, where $h$ is Planck´s constant \cite[see (55.19)]{gottfried}.   As Ramsey noted, \cite[see (IV.34)]{ramsey} the second exponential factor in (1) drives the transitions from one quantum state to another, but is complicated by the non-commutivity of the operators.

From a mathematical viewpoint, this exponential operator in (\ref{gottfriedfunction}) can be simplified as
    \begin{equation}\label{math}
       {\rm e}^{-{\rm i}[({\omega-\Omega}){\bm J}_{z} - \lambda\Omega {\bm J}_{x}]t} \rightarrow {\rm e}^{a{\bf X}+b{\bf Y}}
    \end{equation}
where $a,b$ are scalars and ${\bf X},{\bf Y}$ are operators (or matrices) in a three-element subgroup of the Lie algebra approiate for quantum spins. Then, the separation of ${\rm e}^{a{\bf X}+b{\bf Y}}$ into a simple product of the exponentials ${\rm e}^{\bf X},{\rm e}^{\bf Y}, {\rm e}^{\bf Z}$ of these three operators is not generally allowed by the Baker-Campbell-Hausdorff theorem.

Here we transform this exponential factor  ${\rm e}^{a{\bf X}+b{\bf Y}}$ into a useful form. In Section \ref{cbhd} we  first remind the reader of the Campbell-Baker-Hausdorff-Dynkin (CBHD) theorem, in   which  ${\rm e}^{a{\bf X}+b{\bf Y}}$ is expanded into an infinite series in successive powers of ${\bf X}$ and ${\bf Y}$. Our transformation of this exponential factor into a more physically useful form is presented in Section \ref{theorem}. 
\bigskip

\section{The Campbell-Baker-Hausdorff-Dynkin Formula} \label{cbhd}

According to the CBHD theorem \cite{muger,bonfiglioli}, for any ${\bf X},{\bf Y} \in \mathfrak{g}$, the product of two exponentials of operators or matrices ${\rm e}^{{\bf X}}{\rm e}^{{\bf Y}}$ can be rewritten as ${\rm e}^{H({\bf X},{\bf Y})}$, which is an infinite  series of powers of ${\bf X}$ and ${\bf Y}$,
    \begin{eqnarray}
        {\rm e}^{{\bf X}}{\rm e}^{{\bf Y}}
            & =& {\rm e}^{H({\bf X},{\bf Y})}, \\[0.4em]
        H({\bf X},{\bf Y}) 
           & = &\log{({\rm e}^{{\bf X}}{\rm e}^{{\bf Y}})} \cr
            \qquad\qquad &=& 
            \sum_{k=1}^{\infty}\sum_{m_{1}+n_{1}>0}\cdots \sum_{m_{k}+n_{k}>0}
            \frac{(-1)^{k-1}}{k}\frac{{\bf X}^{m_{1}}{\bf Y}^{n_{1}}\cdots {\bf X}^{m_{k}}{\bf Y}^{n_{k}}}{m_{1}!n_{1}!\cdots m_{k}!n_{k}!}
            \cr
            \qquad\qquad& =& 
            {\bf X}+{\bf Y} + \frac{1}{2}[{\bf X},{\bf Y}] + \frac{1}{12}[{\bf X},[{\bf X},{\bf Y}]] -\frac{1}{12} [{\bf Y},[{\bf X},{\bf Y}]] + \cdots 
            \\[1em] 
            \qquad\qquad& = &\sum_{k=1}^{\infty} \sum_{m_{1}+n_{1}>0} \cdots \sum_{m_{k}+n_{k}>0}
            \frac{(-1)^{k-1}}{k\sum_{i=1}^{k}(m_{i}+n_{i})}
            \frac{1}{m_{1}!n_{1}!\cdots m_{k}!n_{k}!}
            \cr
            \qquad\qquad\quad&&
            \overbrace{[{\bf X},[\cdots, {\bf X}}^{m_{1}},
            \overbrace{[{\bf Y},[\cdots,[{\bf Y}}^{n_{1}},[\cdots
            \overbrace{[{\bf X},[\cdots,[{\bf X}}^{m_{k}},
            \overbrace{[{\bf Y},[\cdots,[{\bf Y},[\cdots]}^{n_{k}}\cdots],
    \end{eqnarray}
where $[{\bf X},{\bf Y}]:={\bf X}{\bf Y}-{\bf Y}{\bf X}$ is the commutator of ${\bf X}$ and ${\bf Y}$ with the understanding that $[{\bf X}]:={\bf X}$.

In this view, the exponential in (\ref{math}) is not allowed to be rewritten in a product form. Instead, one may extend it in an infinite series, as
    \begin{eqnarray}\label{weneed}
        a{\bf X}+b{\bf Y} 
            & =& H(a{\bf{X}},b {\bf{Y}})-\frac{1}{2}[a{\bf {X}},b{\bf {Y}}]-\frac{1}{12}[a{\bf {X}},[a{\bf {X}},b{\bf {Y}}]]+\cdots, \\
        {\rm e}^{a{\bf X}+b{\bf Y}}
            & \neq &{\rm e}^{a{\bf X}}e^{b{\bf Y}}
    \end{eqnarray}

The operator in (\ref{math}) is the exponential of a linear combination of non-commuting spin operators or matrices. To solve for the probability of a transition from a most general state to another most general state is a challenge.

\section{The result} \label{theorem}

We desire to obtain a more useful form of the exponential $e^{a{\bf X}+b{\bf Y}}$ to apply for operators representing physical systems. Our approach is not from the CBHD theorem, but instead from a transformation analogous to a physical rotation for the relevant subgroup of the Lie algebra $\mathfrak{g}$. We confine our Lie algebra to physical systems with three basis operators, and define the form of the unitary transformation needed in order to obtain our results.

\begin{definition}\label{liealgebra}
A set of operators $\{{\bf O}_{\mu}\}=\{{\bf X},{\bf Y},{\bf Z} \}$ is in the Lie algebra $\mathfrak{k}$ if the following 3-cyclic relation is satisfied
    \begin{eqnarray} \label{cyc}
        [{\bf O}_{\mu},{\bf O}_{\nu}]& =& \kappa \epsilon_{\mu\nu\lambda} {\bf O}_{\lambda},
    \end{eqnarray}
where $\epsilon_{\mu\nu\lambda}$ is the Levi-Civita symbol and summation over like Greek subscripts is implied.
\end{definition}

Then, our Lie algebra $\mathfrak{k}$ maintains the Jacobi identity by setting $[{\bf O}_{\mu},{\bf O}_{\mu}]=0$, or explicitly
    \begin{eqnarray} \label{jacobi}
        [{\bf X},[{\bf Y},{\bf Z}]] + [{\bf Y},[{\bf Z},{\bf X}]] + [{\bf Z},[{\bf X},{\bf Y}]]& =& 0,
    \end{eqnarray}
and the spin matrices ${\bm J}_{x},{\bm J}_{y},{\bm J}_{z}$ satisfy Definition \ref{liealgebra} with $\kappa ={\rm i}$ in units of $\hbar$.

\begin{definition}\label{operation}
The transformation of ${\bf O}_{\nu}$ by ${\bf O}_{\mu}$ is defined as
    \begin{equation}
       {\rm  e}^{-p{{\bf O}_{\mu}}} {\bf O}_{\nu} {\rm e}^{p{\bf O}_{\mu}},
    \end{equation}
where $p$ is a scalar. 
\end{definition}

The transformation is analogous to a physical rotation, but we are seeking to apply it to the  exponential of two operators. With the above two definitions, the following theorem demonstrates the existence of a simple product form of the exponentials  ${\rm e}^{\bf X}, {\rm e}^{\bf Y}, {\rm e}^{\bf Z}$ of the elements of the group.

\begin{theorem}
For ${\bf X},{\bf Y},{\bf Z} \in \mathfrak{k}$ and scalar $a,b$, let ${\rm e}^{{\bf U}({\bf X},{\bf Y})}= {\rm e}^{a{\bf X}+b{\bf Y}}$. Then  scalars $p$ and $q$ exist such that 
    \begin{equation}
        {\rm e}^{{\bf U}({\bf X},{\bf Y})} 
            = {\rm e}^{p{\bf Z}}e^{q{\bf X}}{\rm e}^{-p{\bf Z}}.
    \end{equation}
\end{theorem}
 
\noindent{\bf Proof.}
For ${\bf Z},{\bf X}\in \mathfrak{k}$, the derivatives of the transformation of ${\bf X}$ by ${\bf Z}$ with respect to a scalar $p$ are in forms of commutators,
    \begin{eqnarray}
       \frac{d }{d p} \left({\rm e}^{-p{\bf Z}}{\bf X}{\rm e}^{p{\bf Z}}\right) \bigg\vert_{p=0} & = &-[{\bf Z},{\bf X}], \cr
            \qquad\qquad \vdots \cr
            \frac{d^{m} }{d p^{m}} \left({\rm e}^{-p{\bf Z}}{\bf X}{\rm e}^{p{\bf Z}}\right) \bigg\vert_{p=0}& =& (-1)^{m} \big[\overbrace{{\bf Z},[{\bf Z},[\cdots [{\bf Z}}^{m},{\bf X}]\cdots\big].
    \end{eqnarray}
\noindent 
Then, the Taylor expansion of the transformation with respect to $p$ becomes a series of commutators.
    \begin{eqnarray}\label{taylor}
           {\rm  e}^{-p{\bf Z}}{\bf X}{\rm e}^{p{\bf Z}} 
                &=& {\bf X} + (-1)[{\bf Z},{\bf X}]p + \frac{1}{2!}(-1)^2 [{\bf Z},[{\bf Z},{\bf X}]]p^2 + \cdots \cr
                &= &\sum_{m=0}^{\infty} \frac{(-1)^{m}}{m!}[\overbrace{{\bf Z},[{\bf Z},[\cdots[{\bf Z}}^{m},{\bf X}]\cdots]p^{m},
    \end{eqnarray}
where the $m=0$ term in the second line is $\bf{X}$.

Since the operators ${\bf X},{\bf Y},{\bf Z}\in \mathfrak{k}$ satisfy the 3-cyclic relations in (\ref{cyc}), the transformation in (\ref{taylor}) becomes a linear function of ${\bf X}$ and ${\bf Y}$ with trigonometric functions. 
    \begin{eqnarray}
        {\rm e}^{-p{\bf Z}}{\bf X}{\rm e}^{p{\bf Z}} 
            & = &{\bf X}+ {\bf Y} \left[(-1)\kappa p\right]
            + {\bf X}\left[\frac{1}{2!}(-1)\kappa^{2}p^{2}\right] 
            + {\bf Y}\left[\frac{1}{3!}\kappa^{3}p^{3}\right]
            +\cdots\cr
             \cr
             &=&{\bf X}\cos{(\kappa p)} - {\bf Y}\sin{(\kappa p)}.
    \end{eqnarray}
Other analogous transformations are given in Table \ref{Table}.

    \begin{table}[]
    \caption{\label{Table} Transformations of ${\bf X}, {\bf Y}, {\bf Z}$}  
    \noindent
    
        \begin{tabular}{@{}cccccc}
        \hline
        Operation &Equivalence&Operation&Equivalence&Operation &Equivalence\\
        \hline
            ${\rm e}^{-p{\bf X}}{\bf Y}{\rm e}^{p{\bf X}}$ \
            & ${\bf Y}\cos{(\kappa p)} - {\bf Z}\sin{(\kappa p)}$ \
            & ${\rm e}^{-p{\bf X}}{\bf Z}{\rm e}^{p{\bf X}}$ \
            & ${\bf Z}\cos{(\kappa p)} + {\bf Y}\sin{(\kappa p)}$ \
            & ${\rm e}^{-p{\bf X}}{\bf X}{\rm e}^{p{\bf X}}$ \ 
            & ${\bf X}$ \\
            ${\rm e}^{-p{\bf Y}}{\bf Z}{\rm e}^{p{\bf Y}}$ \
            & ${\bf Z}\cos{(\kappa p)} - {\bf X}\sin{(\kappa p)}$ \
            & ${\rm e}^{-p{\bf Y}}{\bf X}{\rm e}^{p{\bf Y}}$ \
            & ${\bf X}\cos{(\kappa p)} + {\bf Z}\sin{(\kappa p)}$\
            & ${\rm e}^{-p{\bf Y}}{\bf Y}{\rm e}^{p{\bf Y}}$ \
            & ${\bf Y}$\\
            ${\rm e}^{-p{\bf Z}}{\bf X}e^{p{\bf Z}}$\
            & ${\bf X}\cos{(\kappa p)} - {\bf Y}\sin{(\kappa p)}$ \
            & ${\rm e}^{-p{\bf Z}}{\bf Y}{\rm e}^{p{\bf Z}}$ \
            & ${\bf Y}\cos{(\kappa p)} + {\bf X}\sin{(\kappa p)}$ \
            & ${\rm e}^{-p{\bf Z}}{\bf Z}{\rm e}^{p{\bf Z}}$ \
            & ${\bf Z}$\\
        \hline
        \end{tabular}\\
    \end{table}
\normalsize

Let ${\bf U}({\bf X},{\bf Y})\in \mathfrak{k}$ be a linear combination of ${\bf X}$ and ${\bf Y}$
    \begin{equation}
        {\bf U}=\{a{\bf X}+b{\bf Y}\vert\ {\bf X},{\bf Y}\in \mathfrak{k},\ a,b\in\mathbb{R} \}.
    \end{equation} 
Then the transformation of ${\bf U}$ by ${\bf Z}$ is
    \begin{eqnarray}\label{generaltrig} 
        {\rm e}^{-p{\bf Z}} {\bf U}({\bf X},{\bf Y}) {\rm e}^{p{\bf Z}}
           & =& {\rm e}^{-p{\bf Z}} \left(a{\bf X}+b{\bf Y}\right){\rm e}^{p{\bf Z}} \cr
           & =& a\left({\rm e}^{-p{\bf Z}} {\bf X}{\rm e}^{p{\bf Z}}\right) + b\left({\rm e}^{-p{\bf Z}} {\bf Y}{\rm e}^{p{\bf Z}}\right) \cr
           & =& {\bf X}\big[a\cos{(\kappa p)}+b\sin{(\kappa p)} \big] 
             +{\bf Y}\big[-a\sin{(\kappa p)}+b\cos{(\kappa p)} \big].
    \end{eqnarray}
The choice of $p$ in (\ref{generaltrig}) is arbitrary. One could first choose $p$ so that the coefficient of ${\bf Y}$ in  (\ref{generaltrig}) vanishes, leading to
    \begin{equation} \label{p}
        p
            =\frac{1}{\kappa}\tan^{-1}(b/a).
    \end{equation}
For general scalar $a$ and $b$, one could write
    \begin{equation}\label{condition1}
        \cos{(\kappa p)}
            = \frac{a}{\sqrt{{{a}}^2+{{b}}^2}},\ 
        \sin{(\kappa p)} 
            = \frac{b}{\sqrt{{{a}}^2+{{b}}^2}.}
    \end{equation}
Then, the transformation of ${\bf U}$ becomes
    \begin{equation}\label{q}
       {\rm  e}^{-p{\bf Z}} {\bf U}({\bf X},{\bf Y}) {\rm e}^{p{\bf Z}} 
            = \sqrt{{{a}}^2+{{b}}^2}{\bf X} \equiv q{\bf X}.
        \smallskip
    \end{equation}
To transform ${\bf U}^{m}$ analogously, inserting the identity matrix $\mathbf{1}={\rm e}^{p{\bf Z}}{\rm e}^{-p{\bf Z}}$ between successive factors of ${\bf U}$ allows one to evaluate ${\rm e}^{-p{\bf Z}}{\bf U}^{m}{\rm e}^{p{\bf Z}}$ precisely. 
    \begin{eqnarray}
     {\rm e}^{-p{\bf Z}} {\bf U}^{m} {\rm e}^{p{\bf Z}}
           & =& \left({\rm e}^{-p{\bf Z}} {\bf U} {\rm e}^{p{\bf Z}}\right)\left({\rm e}^{-p{\bf Z}} {\bf U}^{m-1}{\rm  e}^{p{\bf Z}}\right) \cr
           &=&\left(q{\bf X}\right)^{m},
    \end{eqnarray}
where $q$ is given by Eq.(\ref{q}). Therefore,
    \begin{eqnarray}
        {\rm e}^{-p{\bf Z}}{\rm e}^{{\bf U}({\bf X},{\bf Y})} {\rm e}^{p{\bf Z}}
            &=&{\rm  e}^{-p{\bf Z}} \left( \sum_{m=0}^{\infty} \frac{{\bf U}^{m}}{m!}\right) {\rm e}^{p{\bf Z}}\cr
            &=&{\rm e}^{q{\bf X}}.\cr     \nonumber   
    \end{eqnarray}
    \begin{eqnarray}\label{prooftheorem}
        \therefore\
        {\rm e}^{{\bf U}({\bf X},{\bf Y})} 
            & =&{\rm  e}^{a{\bf X}+b{\bf Y}}\nonumber\\
            & =& {\rm e}^{p{\bf Z}}e^{q{\bf X}}{\rm e}^{-p{\bf Z}}.\hfill \square
    \end{eqnarray}
  
However, the selection of the scalar $p$ is not limited to the above transformation of ${\rm e}^{{\bf U}({\bf X},{\bf Y})}$  in terms of ${\bf X}$. One could also force the coefficient of ${\bf X}$ in  (\ref{generaltrig}) to vanish, leading to $p\rightarrow p'$ given by
    \begin{eqnarray}
        p'
            & =& -\frac{1}{\kappa}\tan^{-1}(a/b)\label{newp1}\\
        {\rm e}^{{\bf U}({\bf X},{\bf Y})}
            & =& {\rm e}^{p'{\bf Z}}{\rm e}^{q{\bf Y}}{\rm e}^{-p'{\bf Z}},\label{newp2}
    \end{eqnarray}
where $q$ is still given by (\ref{q}).
    
One may further extend the theorem for the matrix ${\bf V}({\bf X},{\bf Y},{\bf Z})$, which is a linear combination of ${\bf X},{\bf Y},$ and ${\bf Z}$. 

    \begin{corollary}
    For ${\bf X},{\bf Y},{\bf Z} \in \mathfrak{k}$ and scalar $a,b,c$, let ${{\bf V}({\bf X},{\bf Y},{\bf Z})}= {a{\bf X}+b{\bf Y}+c{\bf Z}}$. Then scalars $p_1,q_1, r$ exist such that  
            \begin{equation}\label{corollary}
               {\rm e}^{{\bf V}({\bf X},{\bf Y},{\bf Z})} ={\rm  e}^{p_1{\bf Z}}{\rm e}^{q_1{\bf Y}}{\rm e}^{r{\bf X}}{\rm e}^{-q_1{\bf Y}}{\rm e}^{-p_1{\bf Z}}.
            \end{equation} 
    \end{corollary}

\noindent {\bf Proof.}
According to the theorem, for ${\bf X},{\bf Y},{\bf Z}\in \mathfrak{k}$, the transformation of ${\bf V}$ with the operator ${\bf Z}$ is 
    \begin{equation}\label{p1}
        {\rm e}^{-p_1{\bf Z}}{{\bf V}}{\rm e}^{p_1{\bf Z}}
            =\sqrt{a^2+b^2}{\bf X}+c{\bf Z},
        \end{equation}
where the coefficient of ${\bf Y}$ has been set equal to zero, and $p_1$ is equal to $p$ in (\ref{p}).  Then,  the second transformation is made with respect to  ${\bf Y}$,
    \begin{eqnarray} \label{choosep1}
        {\rm  e}^{-q_1{\bf Y}}{\rm e}^{-p_1{\bf Z}} {\bf V}{\rm e}^{p_1{\bf Z}}{\rm e}^{q_1{\bf Y}}
            & =& {\rm e}^{-q_1{\bf Y}} \left( \sqrt{{{a}}^2+{{b}}^2}{\bf X} +c{\bf Z} \right){\rm  e}^{q_1{\bf Y}}\nonumber\\
           & =& {\bf X} \left( \sqrt{{{a}}^2+{{b}}^2} \cos{(\kappa q_1)} -c \sin{(\kappa q_1)} \right)\nonumber\\
            &  &\ + {\bf Z} \left( \sqrt{{{a}}^2+{{b}}^2}\sin{(\kappa q_1)} + c\cos{(\kappa q_1)} \right).
    \end{eqnarray}
Selecting $q_1$ so that the coefficient of ${\bf Z}$ vanishes,
\begin{equation}
q_1=-\frac{1}{\kappa}\tan^{-1}{\Big( \frac{c}{\sqrt{a^2+b^2}} \Big)}.
\end{equation}
One may then choose
    \begin{eqnarray}
         \cos(\kappa q_1)&=&\frac{\sqrt{{{a}}^2+{{b}}^2}}{\sqrt{{{a}}^2+{{b}}^2+{{c}}^2}},\nonumber\\ \sin(\kappa q_1)&=&\frac{-c}{\sqrt{{{a}}^2+{{b}}^2+{{c}}^2}}.
     \end{eqnarray}
Then,
\begin{equation}\label{q1}
   {\rm e}^{-q_1{\bf Y}} \left( \sqrt{a^2+b^2}{\bf X} +c{\bf Z} \right) {\rm e}^{q_1{\bf Y}} = \sqrt{{{a}}^2+{{b}}^2+{{c}}^2} {\bf X}\equiv r{\bf X}.
\end{equation}
Therefore, 
    \begin{eqnarray}\label{proofcorollary}
        {\rm e}^{-q_1{\bf Y}}{\rm  e}^{-p_1{\bf Z}}{\rm e}^{{\bf V}({\bf X},{\bf Y},{\bf Z})}{\rm e}^{p_1{\bf Z}} {\rm e}^{q_1{\bf Y}}
            & =&{\rm e}^{r{\bf X}}. \nonumber\\
        \qquad \therefore\
        {\rm e}^{{\bf V}({\bf X},{\bf Y},{\bf Z})} 
            &=&{\rm e}^{p_1{\bf Z}}{\rm e}^{q_1{\bf Y}}{\rm e}^{r{\bf X}}{\rm e}^{-q_1{\bf Y}}{\rm e}^{-p_1{\bf {\bf Z}}}.\square
    \end{eqnarray}

\section{Discussion} \label{discussion}
It is noted that Corollary III.1.1 is not unique. There are actually twelve distinct transformation orderings, resulting from the possible selections of the three basis operators $\{{\bf X}, {\bf Y}, {\bf Z}\}$ and of the three scalars $\{a,b,c\}$. As in (\ref{p1}), by first transforming ${\bf V}({\bf X},{\bf Y},{\bf Z})$ with respect to ${\bf Z}$, one can set the resulting coefficient of either ${\bf X}$ or ${\bf Y}$ equal to zero. Then, by transforming the resultant by the operator, the coefficient of which had been set equal to zero, there could be two additional operator forms analogous to (\ref{proofcorollary}). The remaining eight forms can be obtained from those four forms by cyclic permutations of $\{{\bf X},{\bf Y},{\bf Z}\}$ and $\{a,b,c\}$.

Here we present the other three forms obtained by first transforming ${\bf V}$ with respect to ${\bf Z}$. First, while preserving $p_1$ in the first transformation in (\ref{p1}), one may choose $q'_1$ by forcing the coefficient of ${\bf X}$ to vanish in the second transformation, leading to
    \begin{eqnarray}
        &{\rm e}^{{\bf V}({\bf X},{\bf Y},{\bf Z})}
            = {\rm e}^{p_1{\bf Z}}{\rm e}^{q'_1{\bf Y}}{\rm e}^{r{\bf Z}}{\rm e}^{-q'_1{\bf Y}}{\rm e}^{-p_1{\bf {\bf Z}}},\\
        & q'_1  = \frac{1}{\kappa}\tan^{-1} \big(\frac{\sqrt{a^2+b^2}}{c} \big). \nonumber
    \end{eqnarray}

In the other way, $p_2$ can be chosen for the coefficient of ${\bf X}$ to vanish in the first transformation, similar to ($\ref{newp1}$). Then, in the subsequent transformation with respect to ${\bf X}$, one may force the coefficient of ${\bf Z}$ to vanish, leading to 
    \begin{eqnarray}
      &{\rm e}^{{\bf V}({\bf X},{\bf Y},{\bf Z})} 
            ={\rm e}^{p_2{\bf Z}}{\rm e}^{q_2{\bf X}}{\rm e}^{r{\bf Y}}{\rm e}^{-q_2{\bf X}}{\rm e}^{-p_2{\bf {\bf Z}}},\\
     &\qquad \label{p2}
    p_2=-\frac{1}{\kappa}\tan^{-1}{\big(\frac{a}{b}\big)},\quad
    q_2 = \frac{1}{\kappa}\tan^{-1}{\big(\frac{c}{\sqrt{a^2+b^2}}\big)}.
    \end{eqnarray}
    
Then, with $p_2$ given by (\ref{p2}), another way of transforming ${\bf V}$ is to choose the coefficient of ${\bf Y}$ to vanish in the second transformation. One would obtain 
    \begin{eqnarray}
      &{\rm e}^{{\bf V}({\bf X},{\bf Y},{\bf Z})} 
            ={\rm e}^{p_2{\bf Z}}{\rm e}^{q'_2{\bf X}}{\rm e}^{r{\bf Z}}{\rm e}^{-q'_2{\bf X}}{\rm e}^{-p_2{\bf {\bf Z}}},\\
       &q'_2 = -\frac{1}{\kappa}\tan^{-1}{\big(\frac{\sqrt{a^2+b^2}}{c}\big)}.
    \end{eqnarray}
A summary of the four transformations of ${\rm e}^{\bf V}$ is presented in Table \ref{Table2}, where $r$ is given by (\ref{q1}).

    \begin{table}[h]
    \caption{\label{Table2} Transformations of ${\rm e}^{{\bf V}}$ by First Transformation with Respect to ${\bf Z}$. }
        \noindent
        \begin{tabular}{@{}ccccccc}
        \hline
        $p$&{}&{}&$q$&{}&{}&Equivalence\\
        \hline
            $\quad p_1\quad$
            & $\frac{1}{\kappa}\tan^{-1}{\big(\frac{b}{a}\big)}$ 
            & {\quad}
            & $\quad q_1\quad$
            & $-\frac{1}{\kappa}\tan^{-1}{\big(\frac{c}{\sqrt{a^2+b^2}}\big)}$
            & {\quad}
            & ${\rm e}^{p_1{\bf Z}}{\rm e}^{q_1{\bf Y}}{\rm e}^{r{\bf X}}{\rm e}^{-q_1{\bf Y}}{\rm e}^{-p_1{\bf {\bf Z}}}$\\ 
            $\quad p_1\quad$
            & $\frac{1}{\kappa}\tan^{-1}{\big(\frac{b}{a}\big)}$ 
            & {\quad}
            & $\quad q'_1\quad$
            & $\frac{1}{\kappa}\tan^{-1}{\big(\frac{\sqrt{a^2+b^2}}{c}\big)}$ 
            & {\quad}
            & ${\rm e}^{p_1{\bf Z}}{\rm e}^{q'_1{\bf Y}}{\rm e}^{r{\bf Z}}{\rm e}^{-q'_1{\bf Y}}{\rm e}^{-p_1{\bf {\bf Z}}}$ \\
            $\quad p_2\quad$
            & $-\frac{1}{\kappa}\tan^{-1}{\big(\frac{a}{b}\big)}$ 
            & {\quad}
            & $\quad q_2\quad$
            & $\frac{1}{\kappa}\tan^{-1}{\big(\frac{c}{\sqrt{a^2+b^2}}\big)}$ 
            & {\quad}
            & ${\rm e}^{p_2{\bf Z}}{\rm e}^{q_2{\bf X}}{\rm e}^{r{\bf Y}}{\rm e}^{-q_2{\bf X}}{\rm e}^{-p_2{\bf {\bf Z}}}$ \\
            $\quad p_2\quad$
            & $-\frac{1}{\kappa}\tan^{-1}{\big(\frac{a}{b}\big)}$
            & {\quad}
            & $\quad q'_2\quad$
            & $-\frac{1}{\kappa}\tan^{-1}{\big(\frac{\sqrt{a^2+b^2}}{c}\big)}$ 
            & {\quad}
            & ${\rm e}^{p_2{\bf Z}}{\rm e}^{q'_2{\bf X}}{\rm e}^{r{\bf Z}}{\rm e}^{-q'_2{\bf X}}{\rm e}^{-p_2{\bf {\bf Z}}}$ \\
        \hline
        \end{tabular}
    \end{table}    

Obviously, one could have first transformed ${\bf V}$ with respect to either ${\bf X}$ or ${\bf Y}$, and in each case, there would be {four} choices of the coefficients of the operators to force to vanish, so that there are actually {twelve} expressions for $ {\rm e}^{{\bf V}({\bf X},{\bf Y},{\bf Z})} $ containing only products of exponential factors of a scalar times one of the three operators, ${\bf X}$, ${\bf Y}$, and ${\bf Z}$. The eight remaining transformation forms are obtained by cyclic permutations of $\{{\bf X}, {\bf Y}, {\bf Z}\}$ and $\{a,b,c\}$, and are presented in Table \ref{Table3}. 

\begin{table}[h]
\caption{\label{Table3} Transformation of ${\rm e}^{\bf V}$ by First Transforming with Respect to ${\bf X}$ or ${\bf Y}$.}

\noindent

\begin{tabular}{@{}ccccccc}
\hline
$p$&{}&{}&$q$&{}&{}&Equivalence\\
\hline
    $\quad p_3\quad$
    & $\frac{1}{\kappa}\tan^{-1}{\big(\frac{c}{b}\big)}$  
    & {\quad}
    & $\quad q_3\quad$
    & $-\frac{1}{\kappa}\tan^{-1}{\big(\frac{a}{\sqrt{b^2+c^2}}\big)}$ 
    & {\quad}
    &${\rm e}^{p_3{\bf X}}{\rm e}^{q_{3}{\bf Z}}{\rm e}^{r{\bf Y}}{\rm e}^{-q_{3}{\bf Z}}{\rm e}^{-p_2{\bf {\bf X}}}$\\
    $\quad p_3\quad$
    & $\frac{1}{\kappa}\tan^{-1}{\big(\frac{c}{b}\big)}$ 
    & {\quad}
    & $\quad q'_3\quad$
    & $\frac{1}{\kappa}\tan^{-1}{\big(\frac{\sqrt{b^2+c^2}}{a}\big)}$ 
    & {\quad}
    &${\rm e}^{p_3{\bf X}}{\rm e}^{q'_3{\bf Z}}{\rm e}^{r{\bf X}}{\rm e}^{-q'_3{\bf Z}}{\rm e}^{-p_2{\bf {\bf X}}}$\\
    $\quad p_4\quad$
    & $-\frac{1}{\kappa}\tan^{-1}{\big(\frac{b}{c}\big)}$ 
    & {\quad}
    & $\quad q_4\quad$
    & $\frac{1}{\kappa}\tan^{-1}{\big(\frac{a}{\sqrt{b^2+c^2}}\big)}$ 
    & {\quad}
    & ${\rm e}^{p_4{\bf X}}{\rm e}^{q_{4}{\bf Y}}{\rm e}^{r{\bf Z}}{\rm e}^{-q_{4}{\bf Y}}{\rm e}^{-p_4{\bf {\bf X}}}$ \\
    $\quad p_4\quad$
    & $-\frac{1}{\kappa}\tan^{-1}{\big(\frac{b}{c}\big)}$ 
    & {\quad}
    & $\quad q'_4\quad$
    & $-\frac{1}{\kappa}\tan^{-1}{\big(\frac{\sqrt{b^2+c^2}}{a}\big)}$ 
    & {\quad}
    & ${\rm e}^{p_4{\bf X}}{\rm e}^{q'_4{\bf Y}}{\rm e}^{r{\bf X}}{\rm e}^{-q'_4{\bf Y}}{\rm e}^{-p_4{\bf {\bf X}}}$ \\
    $\quad p_5\quad$
    & $\frac{1}{\kappa}\tan^{-1}{\big(\frac{a}{c}\big)}$ 
    & {\quad}
    & $\quad q_5\quad$
    & $-\frac{1}{\kappa}\tan^{-1}{\big(\frac{b}{\sqrt{c^2+a^2}}\big)}$ 
    & {\quad}
    & ${\rm e}^{p_5{\bf Y}}{\rm e}^{q_{5}{\bf X}}{\rm e}^{r{\bf Z}}{\rm e}^{-q_{5}{\bf X}}{\rm e}^{-p_5{\bf {\bf Y}}}$ \\
    $\quad p_5\quad$
    &$\frac{1}{\kappa}\tan^{-1}{\big(\frac{a}{c}\big)}$ 
    & {\quad}
    & $\quad q'_5\quad$
    & $\frac{1}{\kappa}\tan^{-1}{\big(\frac{\sqrt{c^2+a^2}}{b}\big)}$
    & {\quad}
    & ${\rm e}^{p_5{\bf Y}}{\rm e}^{q'_5{\bf X}}{\rm e}^{r{\bf Y}}{\rm e}^{-q'_5{\bf X}}{\rm e}^{-p_5{\bf {\bf Y}}}$ \\
    $\quad p_6\quad$
    & $-\frac{1}{\kappa}\tan^{-1}{\big(\frac{c}{a}\big)}$ 
    & {\quad}
    & $\quad q_6\quad$
    & $\frac{1}{\kappa}\tan^{-1}{\big(\frac{b}{\sqrt{c^2+a^2}}\big)}$ 
    & {\quad}
    & ${\rm e}^{p_6{\bf Y}}{\rm e}^{q_{6}{\bf Z}}{\rm e}^{r{\bf X}}{\rm e}^{-q_{6}{\bf Z}}{\rm e}^{-p_6{\bf {\bf Y}}}$ \\
    $\quad p_6\quad$
    & $-\frac{1}{\kappa}\tan^{-1}{\big(\frac{c}{a}\big)}$
    & {\quad}
    & $\quad q'_6\quad$
    & $-\frac{1}{\kappa}\tan^{-1}{\big(\frac{\sqrt{c^2+a^2}}{b}\big)}$ 
    & {\quad}
    & ${\rm e}^{p_6{\bf Y}}{\rm e}^{q'_{6}{\bf Z}}{\rm e}^{r{\bf Y}}{\rm e}^{-q'_{6}{\bf Z}}{\rm e}^{-p_6{\bf {\bf Y}}}$ \\
\hline
\end{tabular}
\end{table}

\section{Acknowledgments}\label{acknowledgments}
The authors thank Prof. Joseph Brennan for helpful discussions.  R. A. K. was partially supported by the U. S. Air Force Office of Scientific Research (AFOSR) LRIR \#18RQCOR100, and the AFRL/SFFP Summer Faculty Fellowship Program provided by AFRL/RQ at WPAFB.\vskip10pt
\section{Conflicts of interest statement}
There are no conflicts of interest.\vskip10pt
\section{Data access statement}
The manuscript is self-contained.  There are no data files to access.\vskip10pt
\section{Ethics statement}
No studies on numans, animals, or plants were made.

\section*{References}

\end{document}